\documentclass[12pt,preprint]{aastex}

\newcommand{\OmegaL}{\Omega_{\Lambda}}
\newcommand{\kmsmpc}{{\rm \, km\, s}^{-1}{\rm Mpc}^{-1}}

\begin{document}

\title{Cosmological Constraints from Hubble Parameter versus Redshift Data}
\author{Lado Samushia and Bharat Ratra}
\affil{Department of Physics, Kansas State University, 116 Cardwell Hall, Manhattan, KS 66506}

\begin{abstract}
We use the Simon, Verde, \& Jimenez (2005) determination of the redshift dependence of the Hubble parameter to constrain cosmological parameters in three dark energy cosmological models. We consider the standard $\Lambda$CDM model, the XCDM parameterization of the
dark energy equation 
of state, and a slowly rolling dark energy scalar field with an inverse power-law potential. The constraints are restrictive, consistent
 with those derived from Type Ia supernova redshift-magnitude data, and complement those  from galaxy cluster gas mass fraction versus redshift data.
\end{abstract}

\section{Introduction}

Astrophysical and cosmological data gathered in the last decade strongly support a ``standard'' cosmological model dominated by dark energy. Supernova type Ia (SNIa) redshift-apparent magnitude data show that the Universe is now undergoing accelerated expansion \citep[e.g.,][]{clocchiatti06, astier06, jassal06, conley06, calvo06, carneiro06}. Cosmic microwave background (CMB) data indicate that the Universe has negligible space curvature \citep[e.g.,][]{podariu01b, durrer03,  mukherjee03, page03, spergel06, baccigalupi06}. Many observations indicate that nonrelativistic matter contributes about $30\enspace\%$ of the critical density \citep[and references therein]{chen03b}. These observational facts --- in the context of general relativity --- indicate that we live in a spatially-flat Universe with about $70\enspace\%$ of the total energy density of the Universe today being dark energy, a substance with negative effective pressure responsible for the current accelerated expansion. For reviews see \citet{peebles03}, \citet{carroll04}, \citet{ perivolaropoulos06}, \citet{padmanabhan06}, and \citet{uzan06}, and for discussions of the validity of general relativity on cosmological scales see, e.g., \citet{diazrivera06}, \citet{stabenau06}, \citet{sereno06}, and \citet{caldwell06}.

There are many different dark energy models.\footnote{See \citet{copeland06} for a recent review. For specific models see, e.g., \citet{capozziello05}, \citet{guo06}, \citet{cannata06}, \citet{grande06}, \citet{szydlowski06}, \citet{ nojiri06}, \citet{brax06}, \citet{ calgani06}, and \citet{ guendelman06}.} Here we consider three simple, widely-used ones: standard $\Lambda$CDM, the XCDM parameterization of 
dark energy's equation 
of state, and a slowly rolling dark energy scalar field with an inverse power-law potential ($\phi$CDM). In all three cases we assume that the nonrelativistic matter density is dominated by cold dark matter (CDM). In the $\Lambda$CDM model dark energy is Einstein's cosmological constant $\Lambda$ and can be accounted for in the energy-momentum tensor as a homogeneous fluid with negative pressure $p_{\Lambda}=-\rho_{\Lambda}$ where $\rho_{\Lambda}$ is the cosmological constant energy density \citep{peebles84}. In the $\phi$CDM scenario a scalar field $\phi$ plays the role of dark energy. Here we consider a slowly rolling scalar field with potential energy density $V(\phi)=\kappa m_{\rm p}^2\phi^{-\alpha}$ where $m_{\rm p}$ is Planck's mass and $\kappa$ and $\alpha$ are non-negative constants \citep{peebles88, ratra88}. In the XCDM parameterization dark energy is assumed to be a fluid with pressure $p_{\rm x}=\omega_{\rm x} \rho_{\rm x}$ where $\omega_{\rm x}$ is time-independent and negative but not necessarily equal to $-1$ as in the $\Lambda$CDM model. The XCDM parameterization can be used as an approximation of the $\phi$CDM model in the radiation and matter dominated epochs, but at low redshifts, in the scalar field dominated epoch, a time-independent $\omega_{\rm x}$ is an inaccurate approximation \citep[e.g.,][]{ratra91}. In the $\phi$CDM and XCDM cases we consider a spatially-flat cosmological model while spatial curvature is allowed to be non-zero in the $\Lambda$CDM case. We note that the $\phi$CDM model at $\alpha=0$ and the XCDM parameterization at $\omega_{\rm x}=-1$ are equivalent to a spatially-flat $\Lambda$CDM model with the same matter density.

Besides SNIa and CMB anisotropy, there are many other cosmological tests. Having many tests is important since this allows for consistency checks, and combined together they provide tighter constraints on cosmological parameters. Tests under current discussion include the redshift-angular size test \citep[e.g.,][]{chen03a, podariu03, puetzfeld05, daly05, jackson06}, the galaxy cluster gas mass fraction versus redshift test \citep{sasaki96, pen97, allen04, chen04, kravtsov05, laroque06}, the strong gravitational lensing test \citep{fukugita90, turner90, ratra92, chae04, kochanek04, biesiada06}, the baryonic acoustic oscillation test \citep[e.g.,][]{glazebrook05, angulo05, wang06, zhan06}, and the structure formation test \citep[e.g.,][]{brax05, koivosta05, maor06, bertschinger06, mainini06}. For cosmological constraints from combinations of data sets see, e.g., \citet{wilson06}, \citet{wang06a}, \citet{rahvar06}, \citet{seljak06}, \citet{xia06}, and \citet{rapetti06}.

Here we use a measurement of the Hubble parameter as a function of redshift to derive constraints on cosmological parameters \citep{jimenez02}. \citep[For related techniques see][and references therein.]{shafieloo05, daly05} In our analysis we use the Simon et al. (2005, hereafter SVJ) estimate for the redshift, $z$, dependence of the Hubble parameter,

\begin{equation}
\label{difage}
H(z)=-\frac{1}{1+z}\frac{dz}{dt},
\end{equation}

\noindent where $t$ is time. This estimate is based on differential ages, $dt/dz$, of passively evolving galaxies determined from the Gemini Deep Deep Survey \citep{abraham04} and archival data \citep{dunlop96, spinrad97, treu01, treu02, nolan03}.

SVJ use the estimated $H(z)$ to constrain the dark energy potential and it's redshift dependence. This data has also been used to constrain parameters of holographic dark energy models \citep{yi06}. Here we use the SVJ $H(z)$ data to derive constraints on  cosmological parameters of the $\Lambda$CDM, XCDM, and $\phi$CDM models. In the next section we outline our computation, in $\S\:3$ we present and discuss our results, and we conclude in $\S\:4$.

\section{Computation}

In the ${\Lambda}$CDM model Hubble's parameter is,

\begin{equation}
H(z)=H_0\sqrt{\Omega_m(1+z)^3+\OmegaL +(1-\Omega_m-\OmegaL)(1+z)^2},
\end{equation}

\noindent where $H_0$ is the value of the Hubble constant today and $\Omega_m$ and $\OmegaL$ are the nonrelativistic matter and dark energy density parameters. For the XCDM parameterization in a spatially-flat model,

\begin{equation}
H(z)=H_0\sqrt{\Omega_m(1+z)^3+(1-\Omega_m)(1+z)^{3(1+\omega_{\rm x})}}.
\end{equation}

In the $\phi$CDM model in a spatially-flat Universe the Hubble parameter is,

\begin{equation}
H(z)=H_0\sqrt{\Omega_m(1+z)^3+\Omega_\phi(z)},
\end{equation}

\noindent where the redshift-dependent dark energy scalar field density parameter $\Omega_\phi(z)=[(\dot{\phi})^2+\kappa m_p^2\phi^{-\alpha}]/12$ and an overdot denotes a time derivative. $\Omega_\phi(z)$  has to be evaluated by solving numerically the coupled spatially-homogeneous background equations of motion,

\begin{eqnarray}
\left(\frac{\dot{a}}{a}\right)^2 &=&\frac{8\pi}{3 m_{\rm p}^2}\left[\Omega_m(1+z)^3+\Omega_\phi(z)\right],  \\
& &\ddot{\phi}+3\frac{\dot{a}}{a}\dot{\phi}-\frac{\kappa\alpha}{2}m_p^2\phi^{-(\alpha+1)}=0,
\end{eqnarray}

\noindent where $a(t)$ is the scale factor and $H=\dot{a}/a$. 

To constrain cosmological parameters we use the $H(z)$ data from SVJ. These data, for the redshift range $0.09<z<1.75$, are given in Table 1 and shown in Figure 1.

We determine the best fit values for the model parameters by minimizing

\begin{equation}
\chi^2(H_0,\Omega_m,p)=\sum_{i=1}^9 \frac{[H_{\rm mod}(H_0,\Omega_m,p,z_i)-H_{\rm obs}(z_i)]^2}{\sigma^2(z_i)},
\end{equation}

\noindent where $H_{\rm mod}$ is the predicted value for the Hubble constant in the assumed model, $H_{\rm obs}$ is the observed value, $\sigma$ is the one standard deviation measurement uncertainty, and the summation is over the 9 SVJ data points at redshifts $z_i$. The parameter $p$ describes the dark energy; it is $\OmegaL$ for $\Lambda$CDM, $\omega_{\rm x}$ for XCDM, and $\alpha$ for $\phi$CDM.

$\chi^2(H_0,\Omega_m, p)$ is a function of three parameters. We marginalize the three-dimensional probability distribution function over $H_0$ to get a two-dimensional probability distribution function (likelihood) $L(\Omega_m,p)=\int dH_0P(H_0) e^{-\chi^2(H_0,\Omega_m,p)/2}$. Here $P(H_0)$ is the prior distribution function for Hubble's constant. We consider two Gaussian priors, one with $H_0=73\pm3\enspace\kmsmpc$ \citep[one standard deviation error, from the combination WMAP 3 year estimate,][]{spergel06}, and the other with $H_0=68\pm4\enspace\kmsmpc$  \citep[one standard deviation error, from a median statistics analysis of 461 measurements of $H_0$,][]{gott01, chen03c}. Using $L(\Omega_m,p)$, we define 1, 2, and 3 $\sigma$ contours on the two-dimensional, $(\Omega_m, p)$ parameter space as sets of points with likelihood equal to $e^{-2.30/2}$, $e^{-6.17/2}$, and $e^{-11.8/2}$, respectively,  of the maximum likelihood value.

\section{Results and Discussion}

Figures 2---4 show the $H(z)$ data constraints on the $\Lambda$CDM, XCDM, and $\phi$CDM models. The contours correspond to 1, 2, and 3 $\sigma$ confidence levels from inside to outside. Solid lines are derived using the $H_0=73\pm3\enspace\kmsmpc$ prior while dashed lines correspond to the $H_0=68\pm4\enspace\kmsmpc$ case. 

Figure 2 for the $\Lambda$CDM model shows that the $H(z)$ constraints are most restrictive in a direction approximately orthogonal to lines of constant $0.6\OmegaL -\Omega_m$. In this direction the constraint is as restrictive as that from SNIa redshift-apparent magnitude data \citep[see, e.g.,][Fig. 21]{clocchiatti06}. In the orthogonal direction the $H(z)$ constraint is weaker than that derived using SNIa data. The $H(z)$ constraints complement those derived using galaxy cluster gas mass fraction versus redshift data \citep[see, e.g.,][Fig. 1]{chen04}.

Figure 3 for the XCDM parameterization shows that the $H(z)$ constraints are approximately as constraining as those determined from SNIa redshift-apparent magnitude data \citep[see, e.g.,][Fig. 6]{astier06}, and compliment the constraints derived from galaxy cluster gas mass fraction versus redshift data \citep[see, e.g.,][Fig. 2]{chen04}. At the 2 $\sigma$ confidence level the data favor $\omega_{\rm x}$'s less than $\sim -0.3$ and $\Omega_m$'s less than $\sim 0.5$.

Figure 4 for the $\phi$CDM model shows that the $H(z)$ data constrains $\Omega_m$ much more than $\alpha$. The constraint on the matter density is approximately as tight as the one derived from galaxy cluster gas mass fraction versus redshift data \citep[Fig. 3]{chen04} and from SNIa redshift-apparent magnitude data \citep[Fig. 1]{wilson06}. At the 2 $\sigma$ confidence level the data favor $\Omega_m$'s less than $\sim 0.5$.

The reduced $\chi^2$ values for the best fit models are $\sim 1.8 - 1.9$ for 7 degrees of freedom. This corresponds to a probability $\sim 7 - 8\enspace \%$, a little low, but perhaps not unexpected since this is a first application of the $H(z)$ test.

Our parameter estimates depend on the prior distribution function assumed for $H_0$. This indicates that the $H(z)$ data should be able to constrain $H_0$. If we marginalize the three-dimensional $\Lambda$CDM model likelihood function $L(H_0,\Omega_m,\Omega_\Lambda)=e^{-\chi^2(H_0,\Omega_m,\Omega_\Lambda)/2}$ over $\Omega_m$ and $\Omega_\Lambda$ with uniform priors, we get a probability distribution function for the Hubble constant with best fit value and one standard deviation range of $H_0=61\pm8\enspace\kmsmpc$.
  
\section{Conclusion}

We have used the SVJ Hubble parameter versus redshift data to constrain cosmological parameters of three dark energy models. The constraints are restrictive, and consistent with those determined by using Type Ia supernova redshift-magnitude data. The $H(z)$ data constraints complement those determined from galaxy cluster gas mass fraction versus redshift data. In combination with improved SNIa data \citep[from, e.g., JDEM/SNAP, see, http://snap.lbl.gov/;][and references therein]{podariu01a, crotts05, albert05}, more and better $H(z)$ data will tightly  constrain cosmological parameters. A large amount of $H(z)$ data is expected to become available in the next few years (R. Jimenez, private communication 2006). These include data from the AGN and Galaxy Survey (AGES) and the Atacama Cosmology Telescope (ACT), and by 2009 an order of magnitude increase in $H(z)$ data is anticipated. 

We acknowledge valuable discussions with Raul Jimenez, helpful comments from 
the referee, and support from DOE grant DE-FG03-99ER41093.

\clearpage

\begin{figure} 
\plotone{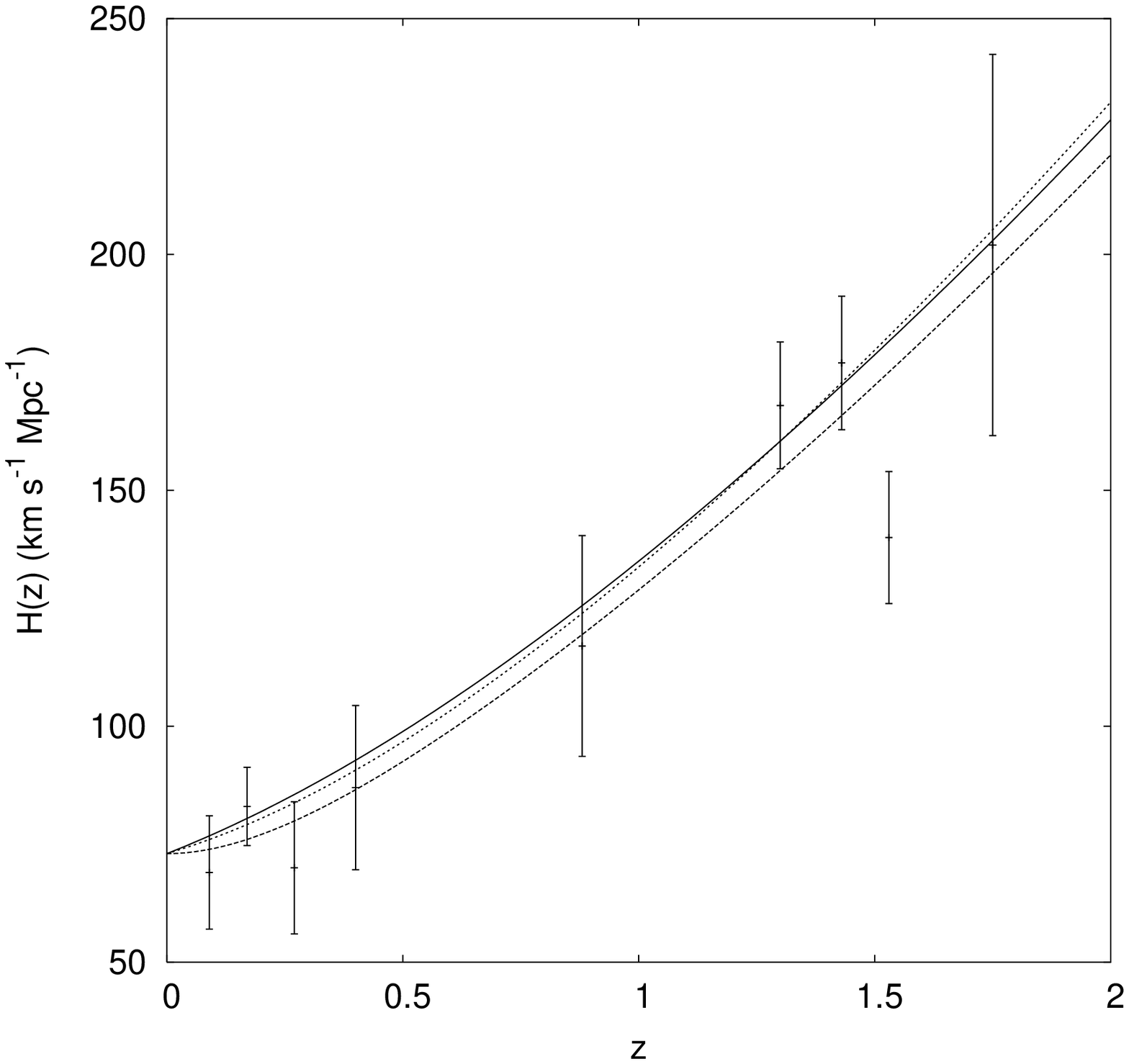}
\caption{SVJ data points with errorbars, and theoretical lines for different dark energy models.
The solid line corresponds to the $\Lambda$CDM model with $\Omega_m=0.34$ and 
$\OmegaL=0.68$, the dotted line corresponds to the spatially-flat XCDM case with $\Omega_m=0.35$ and $\omega_{\rm x}=-1.24$, and the dashed line corresponds to the spatially-flat $\phi$CDM model with $\Omega_m=0.32$ and $\alpha=0.15$. These are best fit models for the case when $H_0=73\enspace\kmsmpc$.}
\end{figure}

\clearpage

\begin{figure} 
\plotone{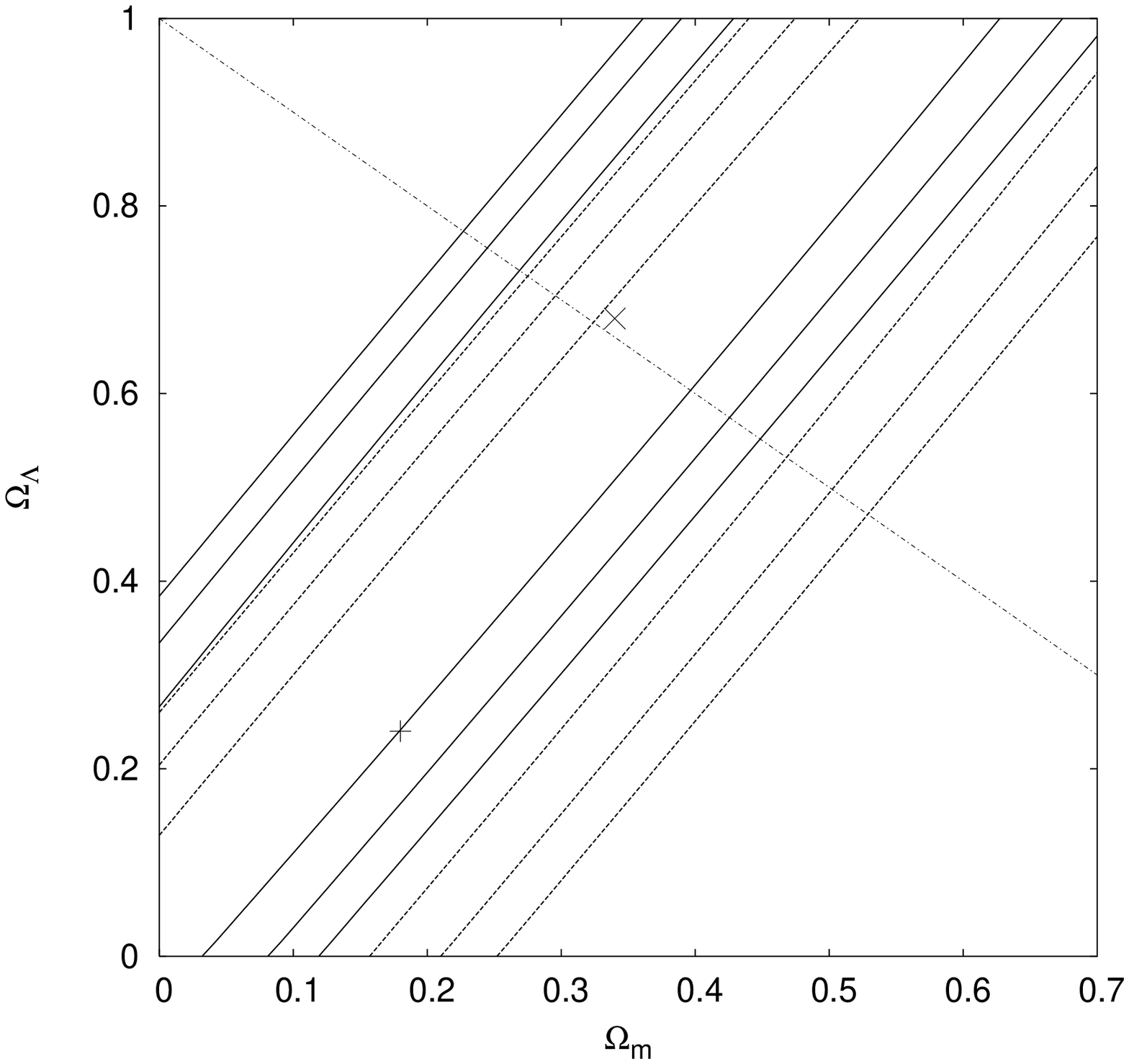}
\caption{1, 2, and 3 $\sigma$ confidence level contours for the $\Lambda$CDM model. Solid lines ($\times$ indicates the maximum likelihood at $\Omega_m=0.34$  and $\OmegaL=0.68$ with reduced $\chi^2=1.80$) correspond to $H_0=73\pm3\enspace\kmsmpc$, while dashed lines ($+$ indicates the maximum likelihood at $\Omega_m=0.18$ and $\OmegaL=0.24$ with reduced $\chi^2=1.88$) correspond to $H_0=68\pm4\enspace\kmsmpc$. The diagonal dot-dashed line indicates spatially-flat $\Lambda$CDM models.}
\end{figure}

\clearpage

\begin{figure} 
\plotone{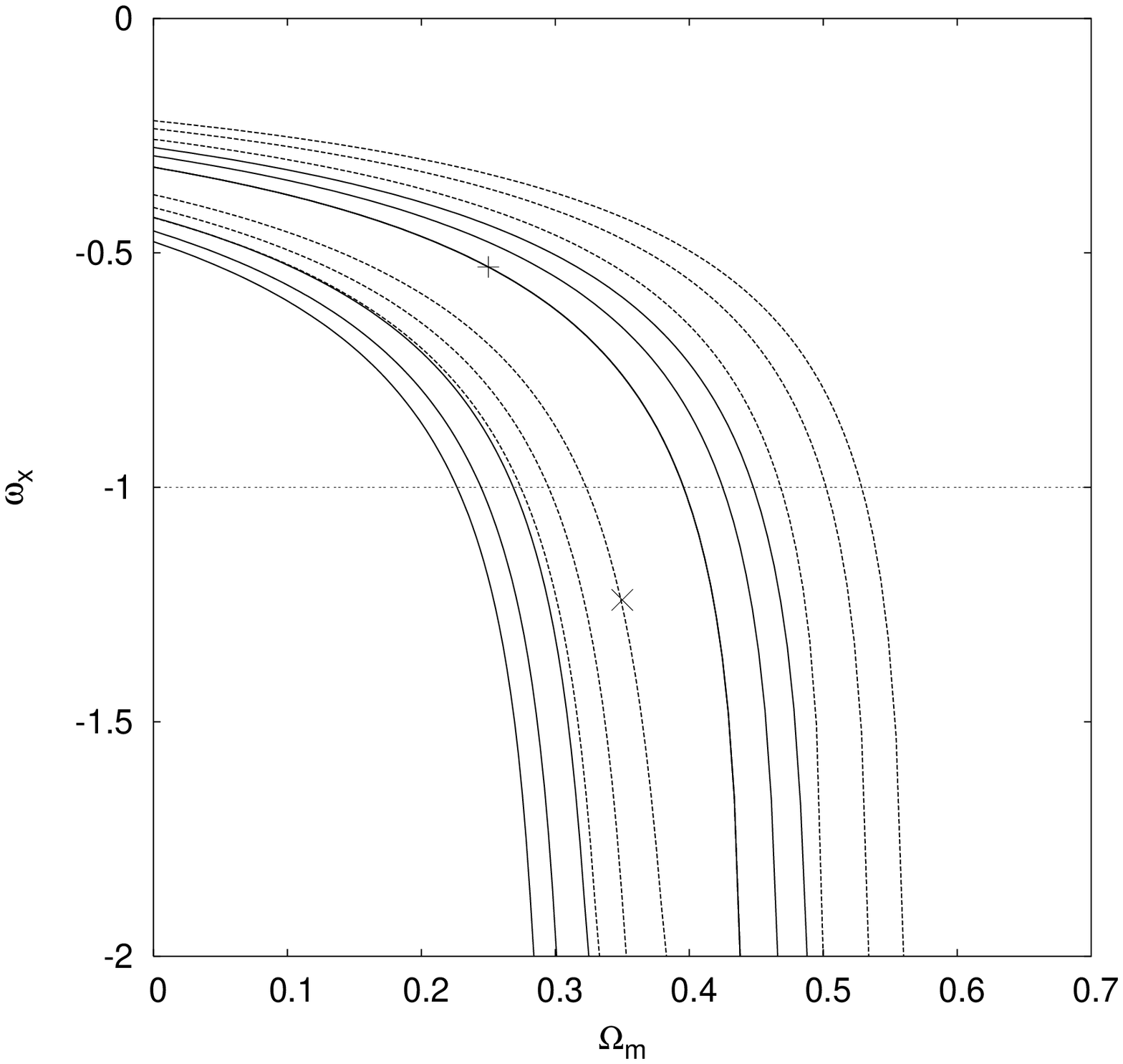}
\caption{1, 2, and 3 $\sigma$ confidence level contours for the spatially-flat XCDM parametrization. Solid lines ($\times$ indicates the maximum likelihood at $\Omega_m=0.35$ and $\omega_{\rm x}=-1.24$ with reduced $\chi^2=1.79$) correspond to $H_0=73\pm3\enspace\kmsmpc$, while dashed lines ($+$ indicates the maximum likelihood at $\Omega_m=0.25$ and $\omega_{\rm x}=-0.53$ with reduced $\chi^2=1.89$) correspond to $H_0=68\pm4\enspace\kmsmpc$. The horizontal dot-dashed line indicates spatially-flat $\Lambda$CDM models.}
\end{figure}

\clearpage

\begin{figure} 
\plotone{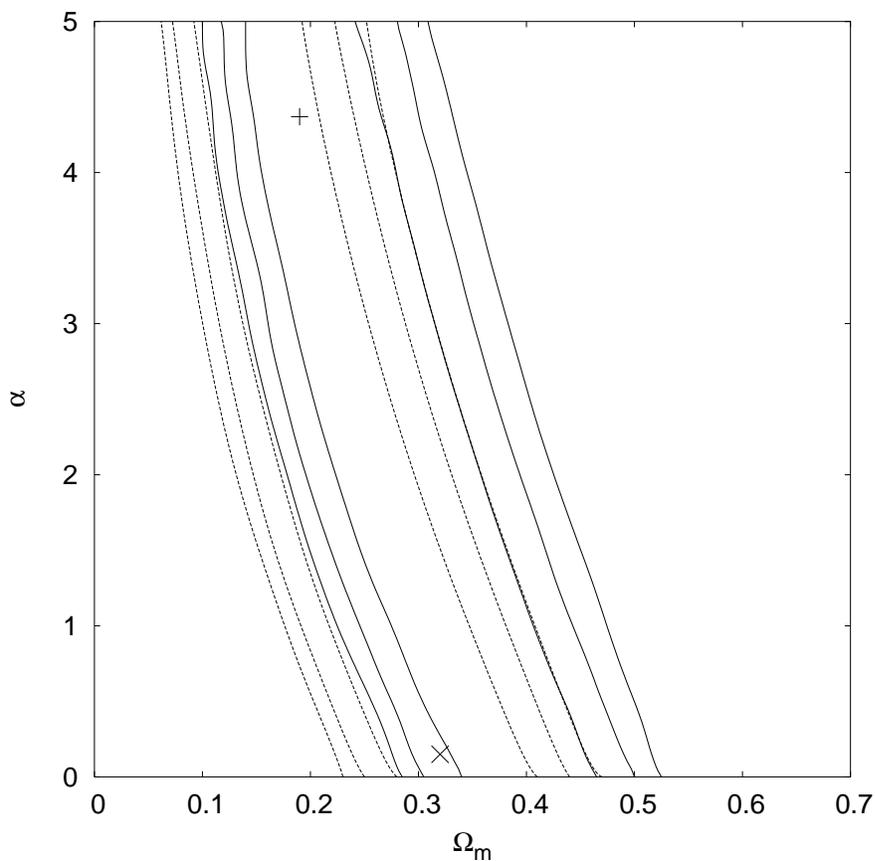}
\caption{1, 2, and 3 $\sigma$ confidence level contours for the spatially-flat $\phi$CDM model. Solid lines ($\times$ indicates the maximum likelihood at $\Omega_m=0.32$ and $\alpha=0.15$ with reduced $\chi^2=1.8$) correspond to $H_0=73\pm3\enspace\kmsmpc$, while dashed lines ($+$ indicates the maximum likelihood at $\Omega_m=0.19$ and $\alpha=4.37$ with reduced $\chi^2=1.89$) correspond to $H_0=68\pm4\enspace\kmsmpc$. The horizontal $\alpha=0$ axis corresponds to spatially-flat $\Lambda$CDM models.}
\end{figure}

\clearpage

\begin{deluxetable}{l l}
\tablewidth{0pt}
\tablecaption{SVJ Hubble Parameter versus Redshift Data}
\tablehead{\colhead{$z$} & \colhead{$H(z)^{\rm a}$} \\ \colhead{} & \colhead{($\kmsmpc$)}}
\startdata
0.09$\quad$&69$\pm$12\\
0.17$\quad$&83$\pm$8.3\\
0.27$\quad$&70$\pm$14\\
0.4 $\quad$&87$\pm$17.4\\
0.88$\quad$&117$\pm$23.4\\
1.3 $\quad$&168$\pm$13.4\\
1.43$\quad$&177$\pm$14.2\\
1.53$\quad$&140$\pm$14\\
1.75$\quad$&202$\pm$40.4\\
\enddata
\tablenotetext{a}{One standard deviation uncertainty.}
\end{deluxetable}


\begin{thebibliography}{}

\bibitem[Abraham et al.(2004)]{abraham04}
  Abraham, R.~G., et al.~2004, AJ, 127, 2455

\bibitem[Albert et al.(2005)]{albert05}
  Albert, J., et al.~2005, astro-ph/0507458

\bibitem[Allen et al.(2004)]{allen04}
  Allen, S.~W., Schmidt, R.~W., Ebeling, H., Fabian, A.~C., \& van Speybroeck, L.~2004, \mnras, 353, 457

\bibitem[Angulo et al.(2005)]{angulo05}
  Angulo, R., et al.~2005, \mnras, 362, L25

\bibitem[Astier et al.(2006)]{astier06}
  Astier, P., et al.~2006, \aap, 447, 31

\bibitem[Baccigalupi \& Acquaviva(2006)]{baccigalupi06}
  Baccigalupi, C., \& Acquaviva, V.~2006, astro-ph/0606069

\bibitem[Bertschinger(2006)]{bertschinger06}
  Bertschinger, E.~2006, astro-ph/0604485

\bibitem[Biesiada(2006)]{biesiada06}
  Biesiada., M.~2006, \prd, 73, 0203006

\bibitem[Brax \& Martin(2006)]{brax06}
  Brax, P., \& Martin, J.~2006, hep-th/0605228

\bibitem[Brax et al.(2005)]{brax05}
  Brax, P., van der Bruck, C., Davis, A., \& Green., A.~2005, Phys. Lett. B., 633, 441

\bibitem[Caldwell \& Grin(2006)]{caldwell06}
  Caldwell, R. R., \& Grin, D.~2006, astro-ph/0606133

\bibitem[Calgani \& Liddle(2006)]{calgani06}
  Calgani, G., \& Liddle, A.~2006, astro-ph/0606003

\bibitem[Calvo \& Maroto(2006)]{calvo06}
  Calvo, G. B., \& Maroto, A. C.~2006, astro-ph/0604409

\bibitem[Cannata \& Kamenshchik(2006)]{cannata06}
  Cannata, F., \& Kamenshchik. A.~2006, gr-qc/0603129

\bibitem[Capozziello et al.(2005)]{capozziello05}
  Capozziello, S., Cardone, V.~F., \& Troisi, A.~2005, astro-ph/0602349

\bibitem[Carneiro et al.(2006)]{carneiro06}
  Carneiro, S., Pigozzo, C., Borges. H. A., \& Alcaniz, J. S.~2006, \prd, 74, 023532

\bibitem[Carroll(2004)]{carroll04}
  Carroll, S.~M.~2004, in Measuring and Modeling the Universe, ed.~W.~L.~Freedman (Cambridge: Cambridge University Press), 235

\bibitem[Chae et al.(2004)]{chae04}
  Chae, K.-H., Chen, G., Ratra, B., \& Lee, D.-W.\ 2004, ApJ, 607, L71

\bibitem[Chen et al.(2003)]{chen03c}
  Chen, G., Gott, J.~R., \& Ratra, B.~2003, PASP, 115, 1269

\bibitem[Chen  \& Ratra(2003a)]{chen03a}
  Chen, G., \& Ratra, B.~2003a, ApJ, 582, 586

\bibitem[Chen \& Ratra(2003b)]{chen03b}
  Chen, G., \& Ratra, B.~2003b  PASP, 115, 1143

\bibitem[Chen \& Ratra(2004)]{chen04}
  Chen, G., \& Ratra, B.~2004,  ApJ, 612, L1

\bibitem[Conley et al.(2006)]{conley06}
  Conley, A., et al.~2006, ApJ, 644, 1 

\bibitem[Copeland et al.(2006)]{copeland06}
  Copeland, E. J., Sami, M., \& Tsujikawa, S.~ 2006, hep-th/0603057

\bibitem[Clocchiatti et al.(2006)]{clocchiatti06}
  Clocchiatti, A., et al. 2006, ApJ, 642, 1

\bibitem[Crotts et al.(2005)]{crotts05}
  Crotts, A., et al.~2005, astro-ph/0507043

\bibitem[Daly \& Djorgovski(2005)]{daly05}
  Daly, R.\ A., \& Djorgovski, S.\ G.\ 2005, astro-ph/0512576

\bibitem[Diaz-Rivera et al.(2006)]{diazrivera06}
  Diaz-Rivera, L. M., Samushia, L., \& Ratra, B.~2006, \prd, 73, 083503
 
\bibitem[Dunlop et al.(1996)]{dunlop96}
  Dunlop, J., et al.~1996, Nature, 381, 581

\bibitem[Durrer et al.(2003)]{durrer03}
  Durrer, R., Novosyadlyj, B., \& Apunevych, S.~2003, \apj, 583, 33

\bibitem[Fukugita et al.(1990)]{fukugita90}
  Fukugita, M., Futamase, T., \& Kasai, M.~1990, MNRAS, 241, 24P

\bibitem[Glazebrook  \& Blake(2005)]{glazebrook05}
  Glazebrook, R., \& Blake, C.~2005, ApJ, 631, 1

\bibitem[Gott et al.(2001)]{gott01}
  Gott, J.~R., Vogeley, M.~S., Podariu, S., \& Ratra, B.~2001, \apj, 549, 1

\bibitem[Grande et al.(2006)]{grande06}
  Grande, J., Sola, J., \& \v Stefan\v ci\' c, F.~2006, gr-qc/0604057

\bibitem[Guo et al.(2006)]{guo06}
  Guo, Z.-H., Ohta, N., \& Zhang, Y.-Z.~2006, astro-ph/0603109

\bibitem[Guendelman \& Kaganovich(2006)]{guendelman06}
  Guendelman, E. I., \& Kaganovich, A. B.~2006, gr-qc/0606017

\bibitem[Jackson \& Jannetta(2006)]{jackson06}
  Jackson, J. C., \& Jannetta, A. L.~2006, astro-ph/0605065

\bibitem[Jassal et al.(2006)]{jassal06}
  Jassal, H.~K., Bagla, J.~S., \& Padmanabhan, T.~2006, astro-ph/0601389

\bibitem[Jimenez \& Loeb(2002)]{jimenez02}
  Jimenez, R., \& Loeb, A.~2005, \apj, 573, 37

\bibitem[Kochanek(2004)]{kochanek04}
  Kochanek, C. S.~2004, astro-ph/0407232

\bibitem[Koivisto \& Mota(2005)]{koivosta05}
  Koivisto, T., \& Mota, D. F.~2005, \prd, 73, 083502

\bibitem[Kravtsov et al.(2005)]{kravtsov05}
  Kravtsov, A.~V., Nagai, D., \& Vikhlinin, A.~A.~2005, ApJ, 625, 588

\bibitem[LaRoque et al.(2006)]{laroque06}
  LaRoque, S.\ J., et al.\ 2006, astro-ph/0604039

\bibitem[Mainini \& Bonometto(2006)]{mainini06}
  Mainini, R., \& Bonometto, S.~2006, \prd, 74, 043504

\bibitem[Maor(2006)]{maor06}
  Maor, I.~2006, astro-ph/0602441

\bibitem[Mukherjee et al.(2003)]{mukherjee03}
  Mukherjee, P. et al.~2003a, Int. J. Mod. Phys. A, 18, 4933

\bibitem[Nojiri et al.(2006)]{nojiri06}
  Nojiri, S., Odintsov. S. D., \& Sami. M.~2006, \prd, 74, 046004

\bibitem[Nolan et al.(2003)]{nolan03}
  Nolan, P. L., et al.~2003, ApJ, 597, 615

\bibitem[Padmanabhan(2006)]{padmanabhan06}
  Padmanabhan, T.~2006, astro-ph/0603114

\bibitem[Page et al.(2003)]{page03}
  Page, L., et al.~2003, ApJS, 148, 233

\bibitem[Peebles(1984)]{peebles84}
  Peebles, P.~J.~E.~1984, \apj, 284, 439

\bibitem[Peebles \& Ratra(1988)]{peebles88}
  Peebles, P.~J.~E., \&\ Ratra, B.~1988, \apj, 325, L17

\bibitem[Peebles \& Ratra(2003)]{peebles03}
  Peebles, P.~J.~E., \&\ Ratra, B.~2003, Rev.~Mod.~Phys., 75, 559

\bibitem[Pen(1997)]{pen97}
  Pen, U.~1997, New Astron., 2, 309

\bibitem[Perivolaropoulos(2006)]{perivolaropoulos06}
  Perivolaropoulos, L.~2006, astro-ph/0601014

\bibitem[Podariu et al.(2003)]{podariu03}
  Podariu, S., Daly, R.\ A., Mory, M., \& Ratra, B.\ 2003 ApJ, 584, 577

\bibitem[Podariu et al.(2001a)]{podariu01a}
  Podariu, S., Nugent, P., \& Ratra, B.~2001a, ApJ, 553, 39

\bibitem[Podariu et al.(2001b)]{podariu01b}
  Podariu, S., Souradeep, T., Gott, J.~R., Ratra, B., \& Vogeley, M.~S.~2001b, ApJ, 559, 9

\bibitem[Puetzfeld et al.(2005)]{puetzfeld05}
  Puetzfeld, D., Pohl, M., \& Zhu, Z.-H.~2005, ApJ, 619, 657

\bibitem[Rahvar \& Movahed(2006)]{rahvar06}
  Rahvar, S., \& Movahed, M. S.~2006, astro-ph/0604206

\bibitem[Rapetti et al.(2006)]{rapetti06}
  Rapetti, D., Allen, S.~W., Amin, M. A., \& Blandford, R. D.~2006, astro-ph/0605683

\bibitem[Ratra(1991)]{ratra91}
  Ratra, B.~1991, \prd, 43, 3802

\bibitem[Ratra \& Peebles(1988)]{ratra88}
  Ratra, B., \& Peebles, P.~J.~E.~1988, \prd, 37, 3406

\bibitem[Ratra  \& Quillen(1992)]{ratra92}
  Ratra, B., \& Quillen, A.~1992, MNRAS, 259, 738

\bibitem[Sasaki(1996)]{sasaki96}
  Sasaki, S.~1996, PASJ, 48, L119

\bibitem[Seljak et al.(2006)]{seljak06}
  Seljak, U., Slosar, A., \& McDonald, P.~2006, astro-ph/0604335

\bibitem[Sereno \& Peacock(2006)]{sereno06}
  Sereno, M. \& Peacock, J.~2006, astro-ph/0605498

\bibitem[Shafieloo et al.(2006)]{shafieloo05}
  Shafieloo, A., Alam, U., Sahni, V., \& Starobinsky, A. A.~2006, MNRAS, 366, 1081 

\bibitem[Simon et al.(2005)]{simon05}
  Simon, J., Verde, L., \& Jimenez, R.~2005, \prd, 71, 123001 (SVJ) 

\bibitem[Spergel et al.(2006)]{spergel06}
  Spergel, D. N., et al.~2006, astro-ph/0603449

\bibitem[Spinrad et al.(1997)]{spinrad97}
  Spinrad, H., et al.~1997, ApJ, 484, 581

\bibitem[Stabenau \& Jain(2006)]{stabenau06}
  Stabenau, H., \& Jain, B.~2006, astro-ph/0604038

\bibitem[Szyd\l{}owski et al.(2006)]{szydlowski06}
  Szyd\l{}owski. M., Kurek, A., \& Krawiec, A.~2006, astro-ph/0604327

\bibitem[Treu et al.(2001)]{treu01}
  Treu, T., et al.~2001, MNRAS, 326, 221

\bibitem[Treu et al.(2002)]{treu02}
  Treu, T., et al.~2002, ApJ, 564, L13

\bibitem[Turner(1990)]{turner90}
  Turner, E.~L.~1990, ApJ, 365, L43

\bibitem[Uzan(2006)]{uzan06}
  Uzan, J.-P.~2006, astro-ph/0605313

\bibitem[Wang(2006)]{wang06}
  Wang, Y.~2006, ApJ, 647, 1

\bibitem[Wang \& Mukherjee(2006)]{wang06a}
  Wang, Y., \& Mukherjee, P~2006, astro-ph/0604051 

\bibitem[Wilson et al.(2006)]{wilson06}
  Wilson. K. M., Chen. G., \& Ratra. B.~2006, astro-ph/0602321

\bibitem[Xia et al.(2006)]{xia06}
  Xia, J.-Q., Zhao, G. B., Li., H., Feng, B, \& Zhang. X.~2006, astro-ph 0605366 

\bibitem[Yi \& Zhang(2006)]{yi06}
  Yi, Z.-L., \& Zhang, T.-J,~2006, astro-ph/0605596

\bibitem[Zhan(2006)]{zhan06}
  Zhan, H.~2006, astro-ph/0605696

\end{thebibliography}
\end{document}